\documentclass[utf8]{frontiersSCNS} 

\usepackage{url,hyperref,lineno,microtype,subcaption, multirow}
\usepackage[onehalfspacing]{setspace}

\usepackage
	{todonotes}

\def\keyFont{\fontsize{8}{11}\helveticabold }
\def\firstAuthorLast{Cécillon \textit{et~al}.} 
\def\Authors{Noé Cécillon, Vincent Labatut, Richard Dufour and Georges Linar{\`e}s}
\def\Address{LIA, Avignon University, France
}
\def\corrAuthor{Vincent Labatut}
\def\corrEmail{\{vincent.labatut\}@univ-avignon.fr}

\begin{document}
\onecolumn
\firstpage{1}

\title[Abusive language detection]{Abusive Language Detection in Online Conversations by Combining Content- and Graph-based Features}

\author[\firstAuthorLast ]{\Authors} 
\address{} 
\correspondance{} 

\extraAuth{}

\maketitle
\begin{abstract}

\section{}

In recent years, online social networks have allowed world-wide users to meet and discuss. As guarantors of these communities, the administrators of these platforms must prevent users from adopting inappropriate behaviors. This verification task, mainly done by humans, is more and more difficult due to the ever growing amount of messages to check. Methods have been proposed to automatize this moderation process, mainly by providing approaches based on the textual content of the exchanged messages. Recent work has also shown that characteristics derived from the structure of conversations, in the form of conversational graphs, can help detecting these abusive messages. In this paper, we propose to take advantage of both sources of information by proposing fusion methods integrating content- and graph-based features. Our experiments on raw chat logs show that the content of the messages, but also of their dynamics within a conversation contain partially complementary information, allowing performance improvements on an abusive message classification task with a final $F$-measure of 93.26\%.

\tiny
 \keyFont{ \section{Keywords:} Automatic abuse detection, Content analysis, Conversational graph, Online conversations, Social networks} 
\end{abstract}

%
%
%
%
%
%
%
%
%

\section{Introduction}
\label{sec:introduction}
Internet widely impacted the way we communicate. Online communities, in particular, have grown to become important places for interpersonal communications. They get more and more attention from companies to advertise their products or from governments interested in monitoring public discourse. Online communities come in various shapes and forms, but they are all exposed to abusive behavior. The definition of what exactly is considered as abuse depends on the community, but generally includes personal attacks, as well as discrimination based on race, religion or sexual orientation.

Abusive behavior is a risk, as it is likely to make important community members leave, therefore endangering the community, and even trigger legal issues in some countries. Moderation consists in detecting users who act abusively, and in taking action against them. Currently this moderation work is mainly a manual process, and since it implies high human and financial costs, companies have a keen interest in its automation. One way of doing so is to consider this task as a classification problem consisting in automatically determining if a user message is abusive or not.

A number of works have tackled this problem, or related ones, in the literature. Most of them focus only on the content of the targeted message to detect abuse or similar properties. For instance,~\cite{Spertus1997} applies this principle to detect hostility,~\cite{Dinakar2011} for cyberbullying, and \cite{Chen2012q} for offensive language. These approaches rely on a mix of standard NLP features and manually crafted application-specific resources (e.g. linguistic rules). We also proposed a content-based method~\citep{Papegnies2017} using a wide array of language features (Bag-of-Words, $tf$-$idf$ scores, sentiment scores). Other approaches are more machine learning intensive, but require larger amounts of data. Recently,~\cite{Wulczyn2017} created three datasets containing individual messages collected from Wikipedia discussion pages, annotated for toxicity, personal attacks and aggression, respectively. They have been leveraged in recent works to train Recursive Neural Network operating on word embeddings and character $n$-gram features~\citep{Pavlopoulos2017, Mishra2018}. However, the quality of these direct content-based approaches is very often related to the training data used to learn abuse detection models. In the case of online social networks, the great variety of users, including very different language registers, spelling mistakes, as well as intentional users obfuscation, makes it almost impossible to have models robust enough to be applied in all cases. \citep{hosseini2017deceiving} have then shown that it is very easy to bypass automatic toxic comment detection systems by making the abusive content difficult to detect (intentional spelling mistakes, uncommon negatives...).

Because the reactions of other users to an abuse case are completely beyond the abuser's control, some authors consider the content of messages occurring \textit{around} the targeted message, instead of focusing only on the targeted message itself. For instance, \citep{Yin2009} use features derived from the sentences neighboring a given message to detect harassment on the Web. \citep{Balci2015} take advantage of user features such as the gender, the number of in-game friends or the number of daily logins to detect abuse in the community of an online game. In our previous work \citep{Papegnies2019}, we proposed a radically different method that completely ignores the textual content of the messages, and relies only on a graph-based modeling of the conversation. This is the only graph-based approach ignoring the linguistic content proposed in the context of abusive messages detection. Our conversational network extraction process is inspired from other works leveraging such graphs for other purposes: chat logs~\citep{Mutton2004} or online forums~\citep{Forestier2011} interaction modeling, user group detection~\citep{Camtepe2004}. Additional references on abusive message detection and conversational network modeling can be found in~\citep{Papegnies2019}.


In this paper, based on the assumption that the interactions between users and the content of the exchanged messages convey different information, we propose a new method to perform abuse detection while leveraging both sources. For this purpose, we take advantage of the content-\citep{Papegnies2017a} and graph-based~\citep{Papegnies2019} methods that we previously developed. We propose three different ways to combine them, and compare their performance on a corpus of chat logs originating from the community of a French multiplayer online game. We then perform a feature study, finding the most informative ones and discussing their role. Our contribution is twofold: the exploration of fusion methods, and more importantly the identification of discriminative features for this problem.

The rest of this article is organized as follows. In Section~\ref{sec:methods}, we describe the methods and strategies used in this work. In Section~\ref{sec:experiments} we present our dataset, the experimental setup we use for this classification task, and the performances we obtained. Finally, we summarize our contributions in Section~\ref{sec:conclusion} and present some perspectives for this work.

\section{Methods}
\label{sec:methods}
\begin{figure}[h!]
    \centering
    \includegraphics[width=\textwidth]{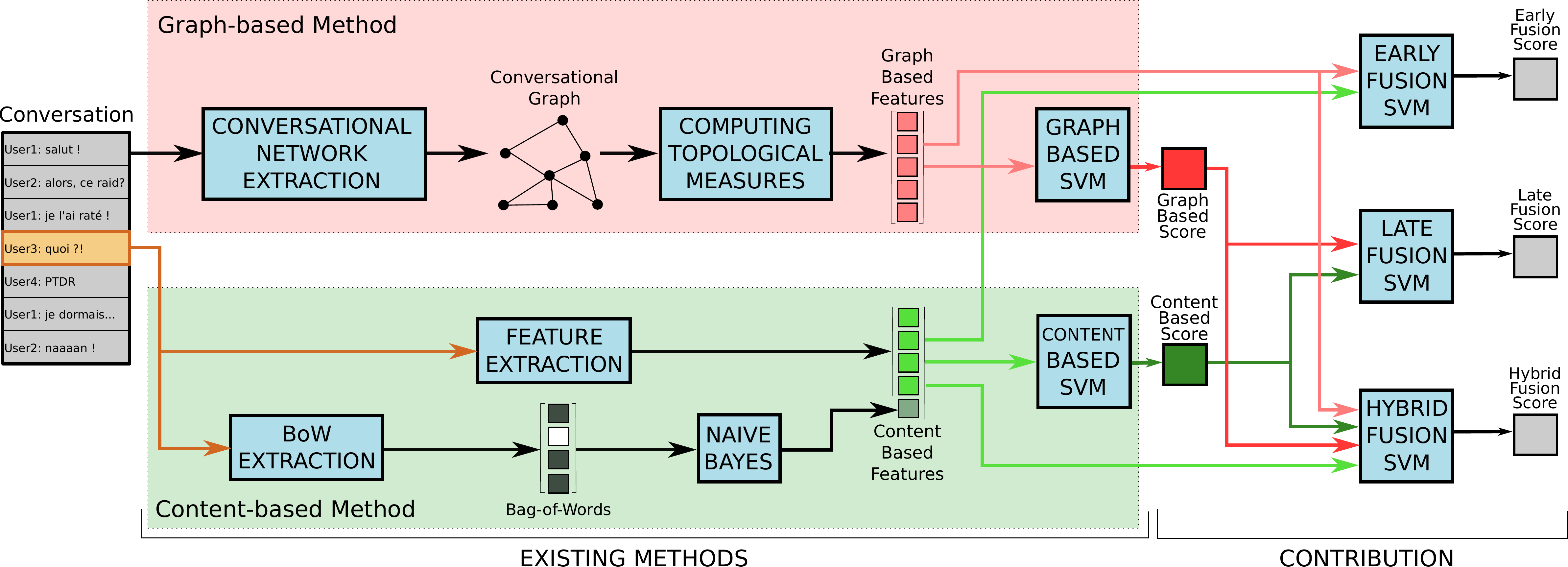}
    \caption{Representation of our processing pipeline. \textit{Existing methods} refers to our previous work described in~\citep{Papegnies2017a} (content-based method) and \citep{Papegnies2019} (graph-based method), whereas the contribution presented in this article appears on the right side (fusion strategies). Figure available at \href{https://doi.org/10.6084/m9.figshare.7442273.v5}{10.6084/m9.figshare.7442273} under CC-BY license.}
    \label{fig:FusionProcess}
\end{figure}

In this section, we summarize the content-based method from~\citep{Papegnies2017a} (Section~\ref{sec:ContentBasedMethod}) and the graph-based method from~\citep{Papegnies2019} (Section~\ref{sec:GraphBasedMethod}). We then present the fusion method proposed in this paper, aiming at taking advantage of both sources of information (Section~\ref{sec:FusionMethod}). Figure~\ref{fig:FusionProcess} shows the whole process, and is discussed through this section.

\subsection{Content-Based Method}
\label{sec:ContentBasedMethod}
This method corresponds to the bottom-left part of Figure~\ref{fig:FusionProcess} (in green). It consists in extracting certain features from the content of each considered message, and to train a Support Vector Machine (SVM) classifier to distinguish abusive (\textit{Abuse} class) and non-abusive (\textit{Non-abuse} class) messages~\citep{Papegnies2017a}. These features are quite standard in Natural Language Processing (NLP), so we only describe them briefly here. 

We use a number of \textit{morphological features}. We use the message length, average word length, and maximal word length, all expressed in number of characters. We count the number of unique characters in the message. We distinguish between six classes of characters (letters, digits, punctuation, spaces, and others) and compute two features for each one: number of occurrences, and proportion of characters in the message. We proceed similarly with capital letters. Abusive messages often contain a lot of copy/paste. To deal with such redundancy, we apply the Lempel–Ziv–Welch (LZW) compression algorithm \citep{Batista2004} to the message and take the ratio of its raw to compress lengths, expressed in characters. Abusive messages also often contain extra-long words, which can be identified by collapsing the message: extra occurrences of letters repeated more than two times consecutively are removed. For instance, “looooooool” would be collapsed to “lool”. We compute the difference between the raw and collapsed message lengths.

We also use \textit{language features}. We count the number of words, unique words and bad words in the message. For the latter, we use a predefined list of insults and symbols considered as abusive, and we also count them in the collapsed message. We compute two overall $tf$--$idf$ scores corresponding to the sums of the standard $tf$--$idf$ scores of each individual word in the message. One is processed relatively to the \textit{Abuse} class, and the other to the \textit{Non-abuse} class. We proceed similarly with the collapsed message. Finally, we lower-case the text and strip punctuation, in order to represent the message as a basic Bag-of-Words (BoW). We then train a Naive Bayes classifier to detect abuse using this sparse binary vector (as represented in the very bottom part of Figure~\ref{fig:FusionProcess}). The output of this simple classifier is then used as an input feature for the SVM classifier.


\subsection{Graph-Based Method}
\label{sec:GraphBasedMethod}
This method corresponds to the top-left part of Figure~\ref{fig:FusionProcess} (in red). It completely ignores the content of the messages, and only focuses on the dynamics of the conversation, based on the interactions between its participants~\citep{Papegnies2019}. It is three-stepped: 1) extracting a conversational graph based on the considered message as well as the messages preceding and/or following it; 2) computing the topological measures of this graph to characterize its structure; and 3) using these values as features to train an SVM to distinguish between abusive and non-abusive messages. The vertices of the graph model the participants of the conversation, whereas its weighted edges represent how intensely they communicate.

\begin{figure}[h!]
    \centering
    \includegraphics[scale=1.1]{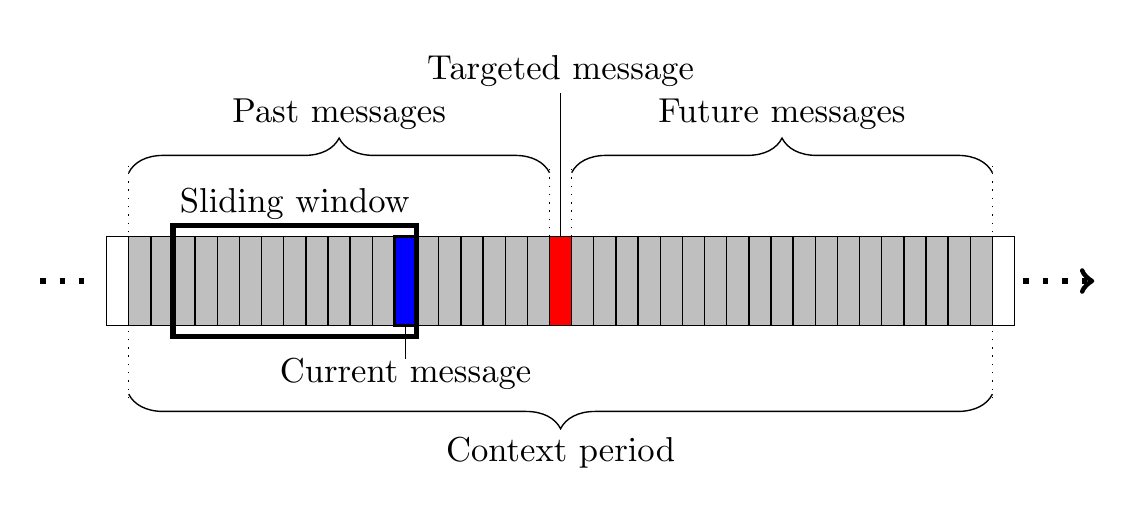}
    \vspace{-0.5cm}
    \caption{Illustration of the main concepts used during network extraction (see text for details). Figure available at \href{https://doi.org/10.6084/m9.figshare.7442273.v5}{10.6084/m9.figshare.7442273} under CC-BY license.}
    \label{fig:ContextPeriod}
\end{figure}

\begin{figure}[!ht]
	\center
	\resizebox{\linewidth}{!}{
		\includegraphics[width=0.333\textwidth]{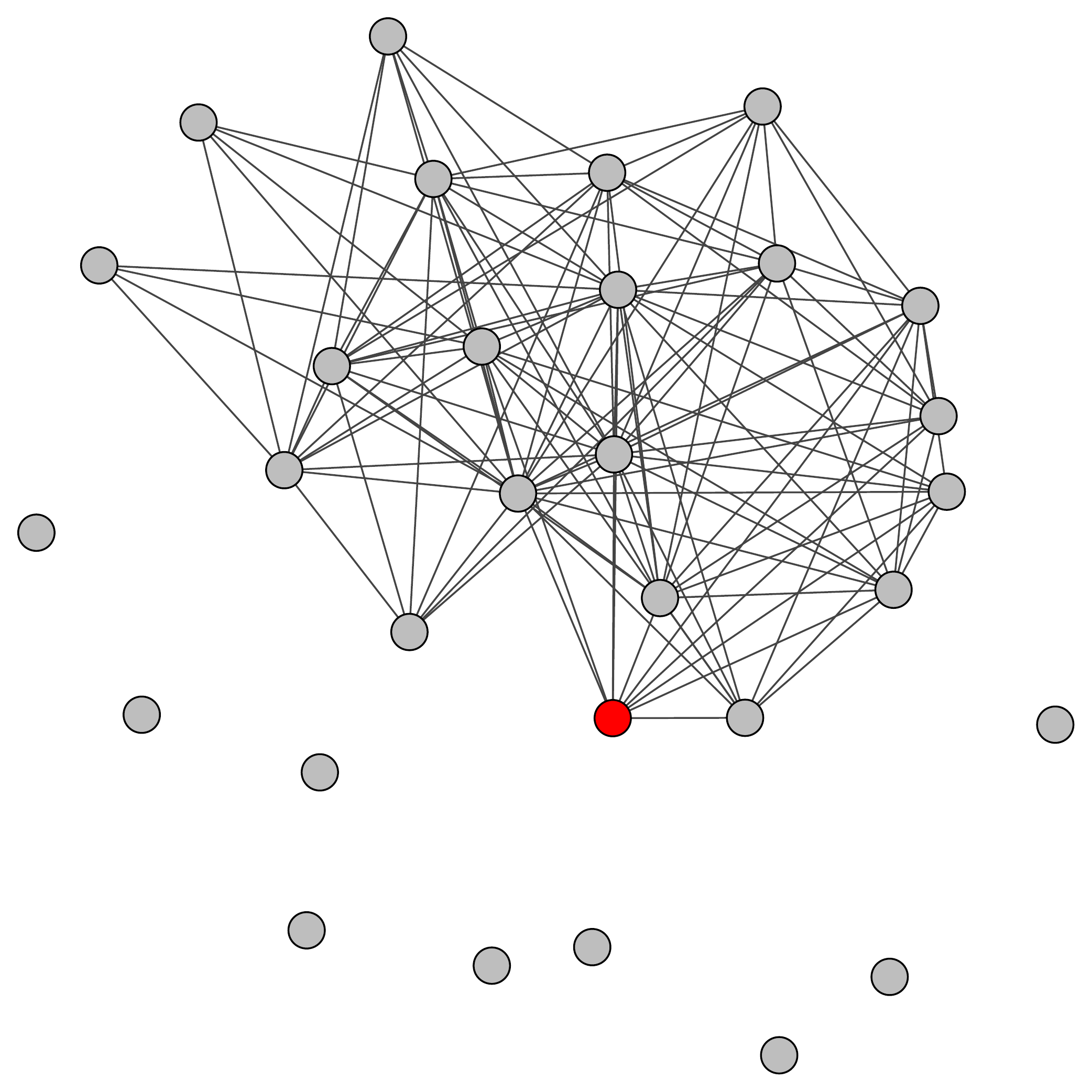}\hfill
		\includegraphics[width=0.333\textwidth]{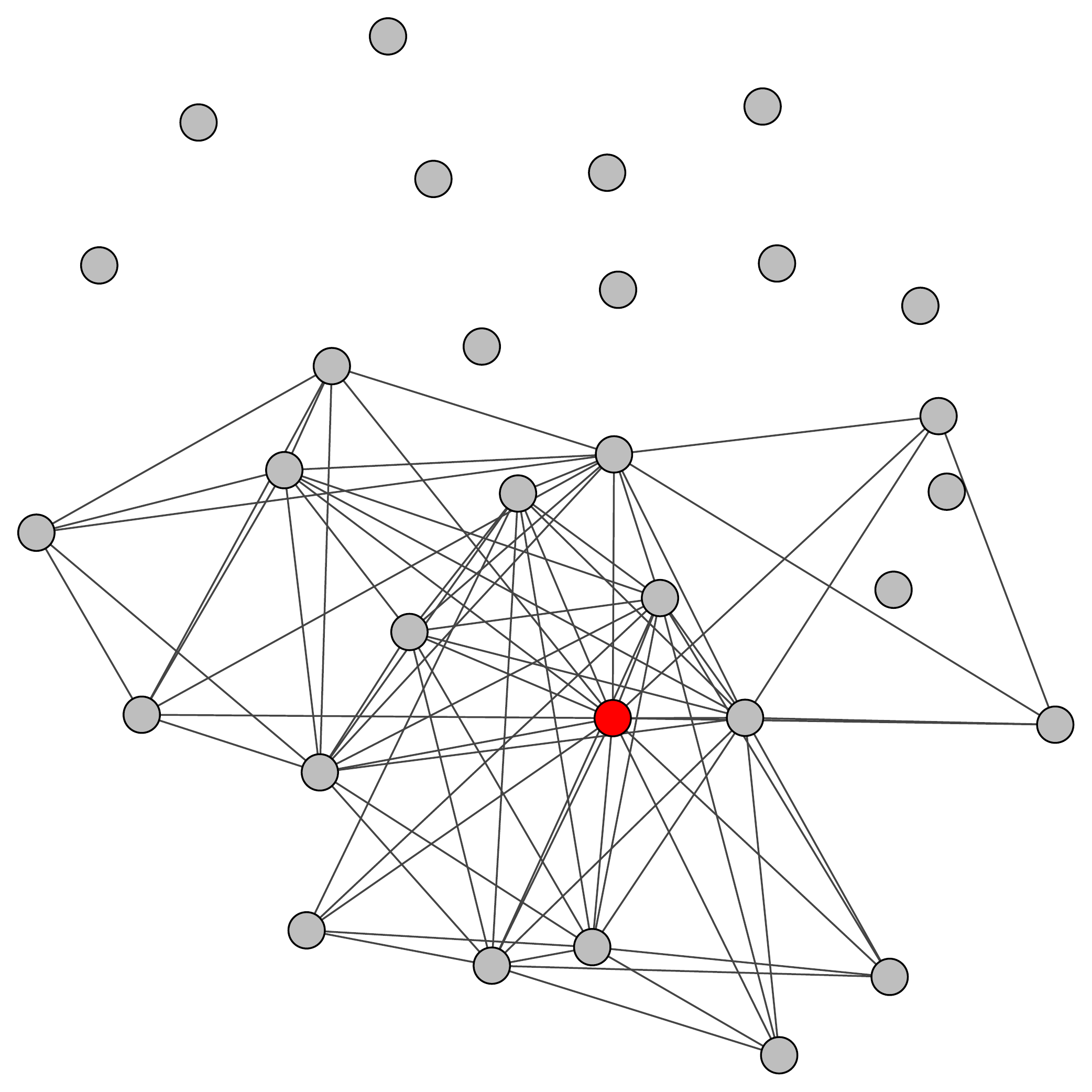}\hfill
		\includegraphics[width=0.333\textwidth]{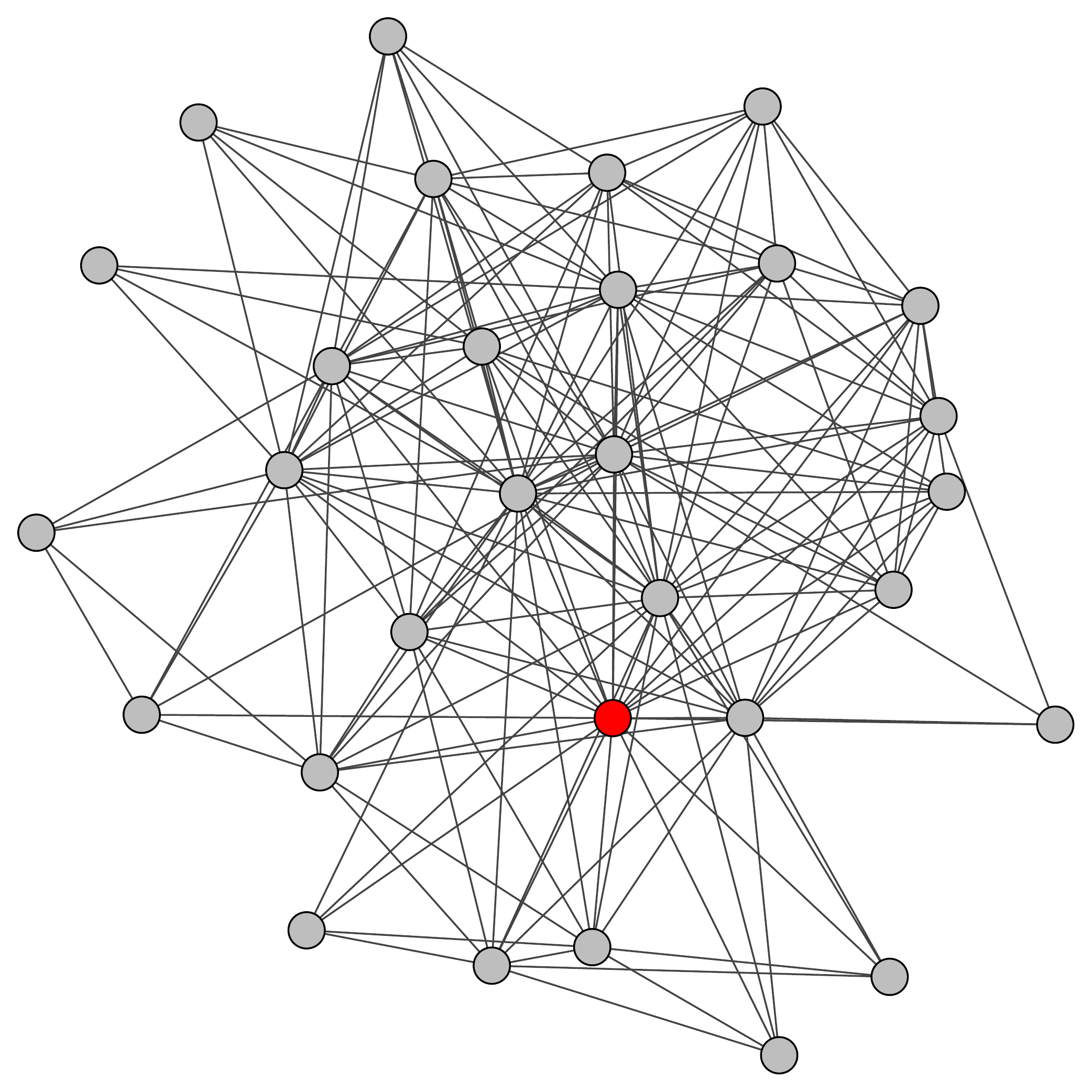}
	}
	\caption{Example of the three types of conversational networks extracted for a given context period: \textit{Before} (left), \textit{After} (center), and \textit{Full} (right). The author of the targeted message is represented in red. Figure available at \href{https://doi.org/10.6084/m9.figshare.7442273.v5}{10.6084/m9.figshare.7442273} under CC-BY license.}
	\label{fig:nets}
\end{figure}

The graph extraction is based on a number of concepts illustrated in Figure~\ref{fig:ContextPeriod}, in which each rectangle represents a message. The extraction process is restricted to a so-called \textit{context period}, i.e. a sub-sequence of messages including the message of interest, itself called \textit{targeted message} and represented in red in Figure~\ref{fig:ContextPeriod}. Each participant posting at least one message during this period is modeled by a vertex in the produced conversational graph. A mobile window is slid over the whole period, one message at a time. At each step, the network is updated either by creating new links, or by updating the weights of existing ones. This \textit{sliding window} has a fixed length expressed in number of messages, which is derived from ergonomic constraints relative to the online conversation platform studied in Section~\ref{sec:experiments}. It allows focusing on a smaller part of the context period. At a given time, the last message of the window (in blue in Figure~\ref{fig:ContextPeriod}) is called \textit{current message} and its author \textit{current author}. The weight update method assumes that the current message is aimed at the authors of the other messages present in the window, and therefore connects the current author to them (or strengthens their weights if the edge already exists). It also takes chronology into account 
by favoring the most recent authors in the window. 
Three different variants of the conversational network are extracted for one given targeted message: the \textit{Before} network is based on the messages posted before the targeted message, the \textit{After} network on those posted after, and the \textit{Full} network on the whole context period. Figure~\ref{fig:nets} shows an example of such networks obtained for a message of the corpus described in Section~\ref{sec:experimental}.

Once the conversational networks have been extracted, they must be described through numeric values in order to feed the SVM classifier. This is done through a selection of standard topological measures allowing to describe a graph in a number of distinct ways, focusing on different scales and scopes. The \textit{scale} denotes the nature of the characterized entity. In this work, the individual vertex and the whole graph are considered. When considering a single vertex, the measure focuses on the \textit{targeted author} (i.e. the author of the targeted message). The \textit{scope} can be either micro-, meso- or macroscopic: it corresponds to the amount of information considered by the measure. For instance, the graph density is microscopic, the modularity is mesoscopic, and the diameter is macroscopic. All these measures are computed for each graph, and allow describing the conversation surrounding the message of interest. The SVM is then trained using these values as features. In this work, we use exactly the same measures as in~\citep{Papegnies2019}.

\subsection{Fusion}
\label{sec:FusionMethod}
We now propose a new method seeking to take advantage of both previously described ones. It is based on the assumption that the content- and graph-based features convey different information. Therefore, they could be complementary, and their combination could improve the classification performance. We experiment with three different fusion strategies, which are represented in the right-hand part of Figure~\ref{fig:FusionProcess}.

The first strategy follows the principle of \textit{Early Fusion}. It consists in constituting a global feature set containing all content- and graph-based features from Sections~\ref{sec:ContentBasedMethod} and~\ref{sec:GraphBasedMethod}, then training a SVM directly using these features. The rationale here is that the classifier has access to the whole raw data, and must determine which part is relevant to the problem at hand.

The second strategy is \textit{Late Fusion}, and we proceed in two steps. First, we apply separately both methods described in Sections~\ref{sec:ContentBasedMethod} and~\ref{sec:GraphBasedMethod}, in order to obtain two scores corresponding to the output probability of each message to be abusive given by the content- and graph-based methods, respectively. Second, we fetch these two scores to a third SVM, trained to determine if a message is abusive or not. This approach relies on the assumption that these scores contain all the information the final classifier needs, and not the noise present in the raw features. 

Finally, the third fusion strategy can be considered as \textit{Hybrid Fusion}, as it seeks to combine both previous proposed ones. We create a feature set containing the content- and graph-based features, like with \textit{Early Fusion}, but also both scores used in \textit{Late Fusion}. This whole set is used to train a new SVM. The idea is to check whether the scores do not convey certain useful information present in the raw features, in which case combining scores and features should lead to better results.

\section{Experiments}
\label{sec:experiments}
In this section, we first describe our dataset and the experimental protocol followed in our experiments (Section~\ref{sec:experimental}). We then present and discuss our results, in terms of classification performance (Sections~\ref{sec:res}) and feature selection (Section~\ref{sec:top}).

\subsection{Experimental protocol}
\label{sec:experimental}
The dataset is the same as in our previous publications~\citep{Papegnies2017a, Papegnies2019}. It is a proprietary database containing 4,029,343 messages in French, exchanged on the in-game chat of \textit{SpaceOrigin}\footnote{\url{https://play.spaceorigin.fr/}}, a Massively Multiplayer Online Role-Playing Game (MMORPG). Among them, 779 have been flagged as being abusive by at least one user in the game, and confirmed as such by a human moderator. They constitute what we call the \textit{Abuse} class. Some inconsistencies in the database prevent us from retrieving the context of certain messages, which we remove from the set. After this cleaning, the \textit{Abuse} class contains 655 messages. In order to keep a balanced dataset, we further extract the same number of messages at random from the ones that have not been flagged as abusive. This constitutes our \textit{Non-abuse} class. Each message, whatever its class, is associated to its surrounding context (i.e. messages posted in the same thread).

The graph extraction method used to produce the graph-based features requires to set certain parameters. We use the values matching the best performance, obtained during the greedy search of the parameter space performed in~\citep{Papegnies2019}. In particular, regarding the two most important parameters (see Section~\ref{sec:GraphBasedMethod}), we fix the \textit{context period} size to 1,350 messages and the \textit{sliding window} length to 10 messages. Implementation-wise, we use the iGraph library~\citep{Csardi2006} to extract the conversational networks and process the corresponding features. We use the Sklearn toolkit~\citep{Pedregosa2011} to get the text-based features. We use the SVM classifier implemented in Sklearn under the name SVC (C-Support Vector Classification). Because of the relatively small dataset, we set-up our experiments using a 10-fold cross-validation. Each fold is balanced between the \textit{Abuse} and \textit{Non-abuse} classes, 70\% of the dataset being used for training and 30\% for testing.

\subsection{Classification Performance}
\label{sec:res}
Table \ref{tab:Perfs} presents the Precision, Recall and $F$-measure scores obtained on the \textit{Abuse} class, for both baselines (\textit{Content-based}~\citep{Papegnies2017a} and \textit{ Graph-based}~\citep{Papegnies2019}) and all three proposed fusion strategies (\textit{Early Fusion}, \textit{Late Fusion} and \textit{Hybrid Fusion}). It also shows the number of features used to perform the classification, the time required to compute the features and perform the cross validation (\textit{Total Runtime}) and to compute one message in average (\textit{Average Runtime}). Note that \textit{Late Fusion} has only 2 direct inputs (content- and graph-based SVMs), but these in turn have their own inputs, which explains the values displayed in the table.

Our first observation is that we get higher $F$-measure values compared to both baselines when performing the fusion, independently from the fusion strategy. This confirms what we expected, i.e. that the information encoded in the interactions between the users differs from the information conveyed by the content of the messages they exchange. Moreover, this shows that both sources are at least partly complementary, since the performance increases when merging them. On a side note, the correlation between the score of the graph- and content-based classifiers is 0.56, which is consistent with these observations.

Next, when comparing the fusion strategies, it appears that \textit{Late Fusion} performs better than the others, with an $F$-measure of 93.26. This is a little bit surprising: we were expecting to get superior results from the \textit{Early Fusion}, which has direct access to a much larger number of \textit{raw} features (488). By comparison, the \textit{Late Fusion} only gets 2 features, which are themselves the outputs of two other classifiers. This means that the \textit{Content-Based} and \textit{Graph-Based} classifiers do a good work in summarizing their inputs, without loosing much of the information necessary to efficiently perform the classification task. Moreover, we assume that the \textit{Early Fusion} classifier struggles to estimate an appropriate model when dealing with such a large number of features, whereas the \textit{Late Fusion} one benefits from the pre-processing performed by its two predecessors, which act as if reducing the dimensionality of the data. 
This seems to be confirmed by the results of the \textit{Hybrid Fusion}, which produces better results than the \textit{Early Fusion}, but is still below the \textit{Late Fusion}. This point could be explored by switching to classification algorithm less sensitive to the number of features. 
Alternatively, when considering the three SVMs used for the \textit{Late Fusion}, one could see a simpler form of a very basic Multilayer Perceptron, in which each neuron has been trained separately (without system-wide backpropagation). This could indicate that using a regular Multilayer Perceptron directly on the raw features could lead to improved results, especially if enough training data is available.

Regarding runtime, the graph-based approach takes more than 8 hours to run for the whole corpus, mainly because of the feature computation step. This is due to the number of features, and to the compute-intensive nature of some of them. The content-based approach is much faster, with a total runtime of less than 1 minute, for the exact opposite reasons. Fusion methods require to compute both content- and graph-based features, so they have the longest runtime.



\begin{table}[!ht]
\centering
\caption{Comparison of the performances obtained with the methods (\textit{Content-based}, \textit{Graph-based}, \textit{Fusion}) and their subsets of \textit{Top Features} (TF). The total runtime is expressed as \textit{h}:\textit{min}:\textit{s}. See text for details.}
  \begin{tabular} { |l|r|r|r|r|r|r| }
    \hline
    \textbf{Method} & \textbf{Number of} & \textbf{Total} & \textbf{Average} & \textbf{Precision} & \textbf{Recall} & \textbf{$F$-measure} \\
     & \textbf{features} & \textbf{Runtime} & \textbf{Runtime} & & & \\
    \hline
    \textbf{Content-Based} & 29 & 0:52 & 0.02s & 78.59 & 83.61 & 81.02 \\
    \textbf{Content-Based TF} & 3 & 0:21 & 0.01s & 75.82 & 82.57 & 79.05 \\ 
    \hline
    \textbf{Graph-Based} & 459 & 8:19:10 & 7.56s & 90.21 & 87.63 &  88.90 \\
    \textbf{Graph-Based TF} & 10 & 14:22 & 0.03s & 88.72 & 84.87 & 86.75 \\ 
    \hline
    \textbf{Early Fusion} & 488 & 8:26:41 & 7.68s & 91.25 & 89.45 & 90.34 \\
    \textbf{Early Fusion TF} & 4 & 11:29 & 0.17s & 89.09 & 87.12 & 88.09 \\ 
    \hline
    \textbf{Late Fusion} & 488 (2) & 8:23:57 & 7.64s & 94.10 & 92.43 & 93.26 \\
    \textbf{Late Fusion TF} & 13 & 15:42 & 0.24s & 91.64 & 89.97 & 90.80 \\ 
    \hline
    \textbf{Hybrid Fusion} & 490 & 8:27:01 & 7.68s & 91.96 & 90.48 & 91.22 \\
    \textbf{Hybrid Fusion TF} & 4 & 16:57 & 0.26s & 90.74 & 89.00 & 89.86 \\ 
    \hline
  \end{tabular}
  \label{tab:Perfs}
\end{table}

\subsection{Feature Study}
\label{sec:top}
We now want to identify the most discriminative features for all three fusion strategies. We apply an iterative method based on the \textit{Sklearn} toolkit, which allows us to fit a linear kernel SVM to the dataset and provide a ranking of the input features reflecting their importance in the classification process. Using this ranking, we identify the least discriminant feature, remove it from the dataset, and train a new model with the remaining features. The impact of this deletion is measured by the performance difference, in terms of $F$-measure. We reiterate this process until only one feature remains. We call \textit{Top Features} (TF) the minimal subset of features allowing to reach $97\%$ of the original performance (when considering the complete feature set).

We apply this process to both baselines and all three fusion strategies. We then perform a classification using only their respective TF. The results are presented in Table~\ref{tab:Perfs}. Note that the \textit{Late Fusion TF} performance is obtained using the scores produced by the SVMs trained on \textit{Content-based TF} and \textit{Graph-based TF}. These are also used as features when computing the TF for \textit{Hybrid Fusion TF} (together with the raw content- and graph-based features). In terms of classification performance, by construction, the methods are ranked exactly like when considering all available features.

The \textit{Top Features} obtained for each method are listed in Table~\ref{tab:TopFeatures}. The last 4 columns precise which variants of the graph-based features are concerned. Indeed, as explained in Section~\ref{sec:GraphBasedMethod}, most of these topological measures can handle/ignore edge weights and/or edge directions, can be vertex- or graph-focused, and can be computed for each of the three types of networks (\textit{Before}, \textit{After} and \textit{Full}).

There are three \textit{Content-Based TF}. The first is the \textit{Naive Bayes} prediction, which is not surprising as it comes from a fully fledged classifier processing BoWs. The second is the \textit{$tf$-$idf$ score} computed over the \textit{Abuse} class, which shows that considering term frequencies indeed improve the classification performance. The third is the \textit{Capital Ratio} (proportion of capital letters in the comment), which is likely to be caused by abusive message tending to be shouted, and therefore written in capitals. The \textit{Graph-Based TF} are discussed in depth in our previous article~\citep{Papegnies2019}. To summarize, the most important features help detecting changes in the direct neighborhood of the targeted author (Coreness, Strength), in the average node centrality at the level of the whole graph in terms of distance (Closeness), and in the general reciprocity of exchanges between users (Reciprocity).

We obtain 4 features for \textit{Early Fusion TF}. One is the \textit{Naive Bayes} feature (content-based), and the other three are topological measures (graph-based features). Two of the latter correspond to the Coreness
of the targeted author, computed for the \textit{Before} and \textit{After} graphs. The third topological measure is his/her Eccentricity. This reflects important changes in the interactions around the targeted author. It is likely caused by angry users piling up on the abusive user after he has posted some inflammatory remark. For \textit{Hybrid Fusion TF}, we also get 4 features, but those include in first place both SVM outputs from the content- and graph-based classifiers. Those are completed by 2 graph-based features, including Strength (also found in the \textit{Graph-based} and \textit{Late Fusion TF}) and Coreness (also found in the \textit{Graph-based}, \textit{Early Fusion} and \textit{Late Fusion TF}).

Besides a better understanding of the dataset and classification process, one interesting use of the TF is that they can allow decreasing the computational cost of the classification. In our case, this is true for all methods: we can retain 97\% of the performance while using only a handful of features instead of hundreds. For instance, with the \textit{Late Fusion TF}, we need only 3\% of the total \textit{Late Fusion} runtime.

\begin{table}[!ht]
	\centering
    \caption{Top features obtained for our 5 methods. The letters in the \textit{Graph} column stand for \textit{Before} (B), \textit{After} (A) and \textit{Full} (F). Those in the \textit{Weights} and \textit{Directions} columns stand for: \textit{Unweighted} or \textit{Undirected} (U), \textit{Weighted} (W), \textit{Directed} (D), \textit{Incoming} (I) and \textit{Outgoing} (O). Those in the \textit{Scale} column mean \textit{Graph-scale} (G) or \textit{Vertex-scale} (N).}
    \begin{tabular}{|r|l|l|l|l|l|}
        \hline
     	\textbf{Method} & \textbf{Top Features} & \textbf{Graph} & \textbf{Weights} & \textbf{Directions} & \textbf{Scale} \\
        \hline
        \multirow{3}{*}{Content-Based} & Naive Bayes & -- & -- & -- & -- \\
		 & $tf$--$idf$ Abuse Score & -- & -- & -- & -- \\
		 & Character Capital Ratio & -- & -- & -- & -- \\
        \hline
		\multirow{10}{*}{Graph-Based} & Coreness Score & F & -- & I & G \\
		 & PageRank Centrality & A & U & D & N \\
		 & Strength Centrality & F & W & O & N \\
		 & Vertex Count & F & -- & -- & G \\
		 & Closeness Centrality & B & W & O & G \\
		 & Closeness Centrality & B & W & O & N \\
		 & Authority Score & B & W & D & G \\
		 & Hub Score & B & U & D & N \\
		 & Reciprocity & A & -- & D & G \\
		 & Closeness Centrality & A & W & U & N \\
        \hline
		\multirow{4}{*}{Early Fusion} & Coreness Score & A & -- & O & G \\
		 & Coreness Score & B & -- & I & G \\
		 & Eccentricity & B & -- & I & G \\
		 & Naive Bayes & -- & -- & -- & -- \\
        \hline
		\multirow{1}{*}{Late Fusion} & \textit{Content-Based} TF $\cup$ \textit{Graph-Based TF} & -- & -- & -- & -- \\
        \hline
		\multirow{4}{*}{Hybrid Fusion} & Graph-based output & -- & -- & -- & -- \\
		 & Content-based output & -- & -- & -- & -- \\
		 & Strength Centrality & A & W & O & N \\
		 & Coreness Score & B & -- & I & G \\
        \hline
  	\end{tabular}
	\label{tab:TopFeatures}
\end{table}

\section{Conclusion and Perspectives}
\label{sec:conclusion}
In this article, we tackle the problem of automatic abuse detection in online communities. We take advantage of the methods that we previously developed to leverage message content~\citep{Papegnies2017} and interactions between users~\citep{Papegnies2019}, and create a new method using both types of information simultaneously. We show that the features extracted from our content- and graph-based approaches are complementary, and that combining them allows to sensibly improve the results up to 93.26 ($F$-measure). One limitation of our method is the computational time required to extract certain features. However, we show that using only a small subset of relevant features allows to dramatically reduce the processing time (down to 3\%) while keeping more than 97\% of the original performance.

Another limitation of our work is the small size of our dataset. We must find some other corpora to test our methods at a much higher scale. However, all the available datasets are composed of isolated messages, when we need threads to make the most of our approach. A solution could be to start from datasets such as the Wikipedia-based corpus proposed by~\cite{Wulczyn2017}, and complete them by reconstructing the original conversations containing the annotated messages. This could also be the opportunity to test our methods on an other language than French. Our content-based method may be impacted by this change, but this should not be the case for the graph-based method, as it is independent from the content (and therefore the language). Besides language, a different online community is likely to behave differently from the one we studied before. In particular, its members could react differently differently to abuse. The Wikipedia dataset would therefore allow assessing how such cultural differences affect our classifiers, and identifying which observations made for Space Origin still apply to Wikipedia.

\bibliographystyle{frontiersinSCNS_ENG_HUMS}
\bibliography{comp}

\end{document}